\documentclass[aps,prb,twocolumn,amsmath,amssymb,floatfix,superscriptaddress]{revtex4-2}
\usepackage{blindtext}
\usepackage{epsfig}
\usepackage{color}
\usepackage{amsmath}
\usepackage{amsfonts}
\usepackage{amssymb}
\usepackage{graphicx}%
\usepackage{url}
\usepackage[breaklinks]{hyperref}
\usepackage[english]{babel}
\usepackage{esvect}
\usepackage[normalem]{ulem}
\usepackage[inline]{enumitem}
\usepackage{xcolor}
\usepackage{xfrac}
\usepackage{dsfont}
\usepackage{bm}
\usepackage{array}
\newcolumntype{C}[1]{>{\centering\let\newline\\\arraybackslash\hspace{0pt}}m{#1}}
\usepackage{footnote}
\usepackage{isomath}
\DeclareMathAlphabet\mathbfcal{OMS}{cmsy}{b}{n}

\hypersetup{colorlinks,citecolor=blue,filecolor=blue,linkcolor=blue,urlcolor=blue}

\bibpunct{[}{]}{,}{n}{}{}

\begin{document}

\title{One-Half Topological Number in Entangled Quantum Physics}
\author{Karyn Le Hur}
\affiliation{CPHT, CNRS, Institut Polytechnique de Paris, Route de Saclay, 91128 Palaiseau, France}

\begin{abstract}
A topological phase can be engineered in quantum physics from the Bloch sphere of a spin-1/2 showing an hedgehog structure as a result of a radial magnetic field. We elaborate on a relation between the formation of an entangled wavefunction at one pole, in a two-spins model, and an interesting pair of one-half topological numbers. Similar to Cooper pairs in superconductors, a Einstein-Podolsky-Rosen pair or Bell state at one pole produces a half flux quantum, which here refers to the halved flux of the Berry curvature on the surface. These 1/2-numbers also refer to the presence of a free Majorana fermion at a pole on each sphere. The topological responses can be measured when driving from north to south and from a circularly polarized field at the poles revealing the quantized or half-quantized nature of the protected transverse currents. We show applications of entangled wavefunctions in band structures, introducing a local marker in momentum space, to characterize the topological response of two-dimensional semimetals in bilayer geometries.

\end{abstract}
\maketitle

\section{Introduction}

Topological phases of matter have attracted a lot of attention these last decades \cite{QiZhang,TIreview}. This gives rise to interesting current flows at the edges protected through a band-gap in the energy spectrum as in the quantum Hall effect \cite{QH}. A topological phase can also be achieved in quantum physics from a spin-1/2 on a Bloch sphere when engineering a Dirac monopole through the action of a radial magnetic field \cite{HH,Roushan,Boulder}. The Berry curvature \cite{Berry} is distributed on the surface of the sphere revealing the presence of a topological charge $q$. This charge gives rise to an integer topological invariant. The hedgehog structure of the spin response on the unit sphere can also be viewed as the formation of a Skyrmion or topological winding number \cite{Skyrmion1,Skyrmion2}.  Geometrical properties are measured when driving from north to south \cite{HH,Roushan,Boulder} and can also be revealed through the application of a time-dependent circularly polarized field \cite{KLHlight}. The sphere model also finds applications in energy through a dynamo effect \cite{dynamo1,dynamo2}. In Ref. \cite{HH}, we have introduced a pair of one-half topological numbers from the Bloch sphere as a result of entanglement formation. Coupling two spins in curved space reveals a stable one-half topological number on each sphere, within the phase diagram, when driving from north to south \cite{HH}. 
In this Article, through Eqs. (\ref{identity}) and (\ref{charge}), we show how the number $q\frac{1}{2}$ can be justified from an Einstein-Podolsky-Rosen (EPR) pair or Bell state \cite{AlainAspect} at one pole and a pure state at the other pole. 

We relate entanglement properties and the one-half number through bipartite fluctuations \cite{bipartitefluctuations}. The one-half number can be measured in transport and also from the response to a circularly polarized field acting locally on one sphere. These half numbers can also be understood in terms of Majorana fermions \cite{Majorana,Majoranareturns}, which are their own antiparticles, each sphere revealing one free Majorana fermion at the pole where the EPR pair forms. The two-spheres' model can be implemented to realize a pair of half-Skyrmions (merons), with only half of the surface radiating the Berry curvature. The quest of Skyrmions and fractional-Skyrmions has engendered a lot of curiosity in physics e.g. related to the Yang-Mills equation \cite{Actor} towards recent applications in quantum Hall systems \cite{Sondhi,Shayegan,Potemski}, three-dimensional topological insulators \cite{QiZhang,Xu} and magnetic materials \cite{Jena,Fert,Nagaosa}. In Ref. \cite{HH}, we introduced a topological semimetal on the honeycomb lattice, in a bilayer system, showing a phase diagram similar to the model of two spheres. Through Eqs. (\ref{wavefunctionK'}) and (\ref{equationK'}), we show that a local $q\frac{1}{2}$ topological marker can be introduced, revealing the presence of a pair of half-Skyrmions and entanglement properties at the Dirac point related to a nodal ring semimetal.

The class of models we address takes the form of two interacting spins-1/2 on a Bloch sphere \cite{HH}
\begin{equation}
\label{model}
H = - {\bf d}_1.\mathbfit{\sigma}_1 - {\bf d}_2.\mathbfit{\sigma}_2 + rf(\theta)\sigma_{1z}\sigma_{2z}
\end{equation}
with
\begin{equation}
{\bf d}_i =  (d\sin\theta\cos\phi, d\sin\theta\sin\phi, d\cos\theta +M_i).
\end{equation}
We implicitly assume inversion symmetry between the two spins such that $M_1=M_2=M>0$, that will be at the heart of the protection of the results.
The function $f(\theta)$ can be set to unity even though conclusions remain identical for a large class of functions $f(\theta)$. This model is realizable in mesoscopic and atomic physics \cite{Roushan}. 

\section{General Formalism}

Here, we introduce the geometrical formalism for one sphere with $r=0$. The magnetic field ${\bf d}_i$ acting on a Bloch sphere associated to each spin-$\frac{1}{2}$ encodes the presence of a Dirac monopole or Skyrmion in a sphere (hedgehog). A topological phase with one magnetic charge $q$ in each sphere exists as long as $M<d$. The parameter $M$ corresponding to a uniform magnetic field will allow for phase transitions. The two-particles wavefunction of the model (\ref{model}) takes the form
\begin{equation}
\label{Ket}
|\psi\rangle = \sum_{kl} o_{kl}(\theta) |\Phi_k(\phi)\rangle_1 |\Phi_l(\phi)\rangle_2.
\end{equation}
Here, $|\Phi_+\rangle$ and $|\Phi_-\rangle$ are the usual 2-components spinors with norm unity associated to eigenstates $|+\rangle_z$ and $|-\rangle_z$ with ${\mathbfit\sigma}_{jz}=\pm 1$, which then form the Hilbert space basis. The dependence on the azimuthal angle $\phi$, which acts as a gauge choice in the wavefunction, is included through $|\Phi_k(\phi)\rangle_i$.  The amplitudes $o_{kl}$ depend on the polar angle $\theta$ only for this model which can be understood from the fact that at the poles of the sphere all $\phi$ angles are equivalent and from the rotational symmetry perpendicular to the $z$-axis. For $M<d$ and $r=0$, the ground state evolves from $|\psi(0)\rangle= |\Phi_+(\phi)\rangle_1 \otimes |\Phi_+(\phi)\rangle_2$ onto $|\psi(\pi)\rangle_{r=0} = |\Phi_-(\phi)\rangle_1 \otimes |\Phi_-(\phi)\rangle_2$ and follows the direction of the ${\bf d}_j$ vector. The subscript $r=0$ will be introduced hereafter to specify that a quantity is evaluated at $r=0$, where it needs to be specified. Geometrical properties are  encoded through the Berry gauge potential such that it acts on each spin (sphere), e.g. $A_{1\phi}=-i\langle \psi| \partial_{1\phi}\otimes\mathbb{I}|\psi\rangle$ and the Berry curvature reads $F^1_{\theta\phi} = \partial_{\theta} A_{1\phi}$. We have similar definitions for $A_{2\phi} = -i\langle \psi| \mathbb{I}\otimes \partial_{\theta} A_{2\phi}|\psi\rangle$ and $F^2_{\theta\phi} = \partial_{\theta} A_{2\phi}$. When varying $r$, through this Article, the wavefunction at $\theta=0$ will remain identical such that $A_{j\phi,r=0}(0)=A_{j\phi}(0)$. The functions $A_{j\phi}$ depend on $\theta$ only and $A_{j\theta}=0$.

For $M=0$ and $r=0$, the ground state $|\psi\rangle=|\psi_G\rangle=|\psi_{1G}\rangle\otimes |\psi_{2G}\rangle$ corresponds to an energy $-2d$, and
the amplitudes $o_{kl}$ satisfy simple relations $o_{++}(\theta)=\cos^2\frac{\theta}{2}$, $o_{--}(\theta)=\sin^2\frac{\theta}{2}$, $o_{+-}(\theta)=o_{-+}(\theta)=\frac{\sin\theta}{2}$. 
For $r=0$ and $M=0$, this gives rise to the simple identity
\begin{equation}
\label{functionA}
A_{j\phi}(\theta) = \cos^2\frac{\theta}{2} A_{j\phi}(0) + \sin^2\frac{\theta}{2}A_{j\phi}(\pi).
\end{equation}
For the symmetric gauge choice $|\psi_{jG}\rangle = \cos\frac{\theta}{2}e^{-i\frac{\phi}{2}}|\Phi_+\rangle_j + \sin\frac{\theta}{2}e^{i\frac{\phi}{2}}|\Phi_-\rangle_j$, $A_{j\phi}(\theta)=-\frac{q\cos\theta}{2}$ leading to the gauge-invariant form $F_{\theta\phi}^j(\theta) = \frac{q \sin\theta}{2}$, with $q=1$ the topological charge. The topological invariant measuring the presence of a Dirac monopole in each sphere can be
introduced as in cartesian coordinates \cite{Roushan}, and takes the two equivalent forms \cite{HH,KLHlight}
\begin{eqnarray}
\label{marker}
C_j &=& \frac{1}{2\pi} \int_{0}^{\pi}\int_0^{2\pi} F^j_{\theta\phi}(\theta) d\theta d\phi \\ \nonumber
 &=& A_{j\phi}(\pi) - A_{j\phi}(0) = q.
\end{eqnarray} 
Due to the inversion symmetry between the two spheres, the charge is identical in each sphere such that $q=q_1=q_2$.
This is equivalently measurable locally from the Berry phases at the poles of the Poincar\' e or Riemann sphere through the function $A_{j\phi}$ defined smoothly on the whole surface \cite{HH,KLHlight}. The function $A_{j\phi}(\theta)$ is measurable through $A_{j\phi}=-\frac{q}{2}\langle \psi_{jG}| \sigma_{jz}(\theta) |\psi_{jG}\rangle+A_{j\phi}(Ref)$ and $A_{j\phi}(Ref)=0$ for the symmetric gauge. In this Article, we also report interesting relations between quantum metric or quantum distance \cite{Carollo}, geometrical properties and responses to a circularly polarized field \cite{KLHlight}. Activating the azimuthal angle $\phi\rightarrow \phi'$ on a sphere $j$,  such that $M_1=M_2=0$ and $r=0$, this leads to 
\begin{equation}
\hskip -0.2cm
\label{metric}
|\langle \psi_{jG}(\theta,\phi) | \psi_{jG}(\theta,\phi')\rangle|^2 = \alpha_j(\theta) + 2\cos(\phi'-\phi)(F^j_{\theta\phi})^2,
\end{equation}
where the function $\alpha_j(\theta)=\cos^4\frac{\theta}{2}+\sin^4\frac{\theta}{2}$ also measures the response to circularly polarized light \cite{KLHlight}. 
As long as $0<M<d$, the topological charge sits in the core of each sphere and the response takes an identical form modulo a redefinition of the polar angle $\theta\rightarrow \tilde{\theta}$ such that $\tan\tilde{\theta}=\sin\theta/(\cos\theta+\frac{M}{d})$ with $\tilde{\theta}\in [0;\pi]$.

When $M\rightarrow d$, then $\tilde{\theta}\in [0;\frac{\pi}{2}]$. It is similar as if the surface is reduced to an hemisphere which corresponds to a half topological number $C_j=\frac{1}{2}$ or a half Skyrmion. 
From Eq. (\ref{functionA}), this leads to $C_j= A_{j\phi}(\tilde{\theta}=\frac{\pi}{2})- A_{j\phi}(\tilde{\theta}=0)= \frac{q}{2}$. For $M>d$, $\theta=\pi$ leads to $\tilde{\theta}\rightarrow 0$ such that $C_j=0$. It is as if the topological charge leaks out from the sphere. The jump of the topological number at the quantum phase transition is observed in Ref. \cite{Roushan}. It is also important to mention measures of the quantum metric tensor related to Eq. (\ref{metric}) in superconducting circuits \cite{Tan}. 

The topological characterization at the poles on the sphere is elegant to describe topological lattice models such as the Haldane model \cite{Haldanemodel} on the honeycomb lattice \cite{Klein} and topological p-wave superconducting wires \cite{pwavewire}.

\section{Fractional Topological Number from the poles}

 For two interacting spheres, as long as $r<d-M$, the physics remains identical with $C_j=1$ while for $r>d+M$, the two spheres are described with a topological number $C_j=0$ due to the fact that the states at the two poles become identical, $A_{j\phi}(0)=A_{j\phi}(\pi)$. A fractional phase with $C_j=\frac{1}{2}$ develops on a line when adjusting $r$ such that $d-M<r<d+M$ which then spreads when adding an $r_{xy}(\sigma_{1x}\sigma_{2x}+\sigma_{1y}\sigma_{2y})$ interaction. In Ref. \cite{HH}, the proofs showing the existence of the fractional topological number were derived developing the geometrical approach through Stokes' theorem, and also through a numerical analysis when varying the polar angle linearly in time $\theta=vt$ for various forms of $f(\theta)$ functions. Here, we derive local identities showing the existence of such fractional topological numbers from the forms of the two-particles wavefunctions at the poles \cite{HH}, $|\psi(0)\rangle=|\Phi_+\rangle_1\otimes |\Phi_+\rangle_2$ and $|\psi(\pi)\rangle=\frac{1}{\sqrt{2}}(|\Phi_+\rangle_1\otimes |\Phi_-\rangle_2 + |\Phi_-\rangle_1\otimes |\Phi_+\rangle_2)$. It is interesting to mention that at $\theta=\pi$ the two states $|\Phi_+\rangle_1\otimes |\Phi_-\rangle_2$ and $|\Phi_-\rangle_1\otimes |\Phi_+\rangle_2$ are degenerate and the Einstein-Podolsky-Rosen (EPR) pair or Bell state occurs as a result of perturbation theory giving rise to a term $-\lambda\sigma_{1x}\sigma_{2x}$, with $\lambda\sim \frac{d^2\sin^2\theta}{r}$ if $r\gg d-M$, in the Hamiltonian favoring $|\psi(\pi)\rangle$ compared to the singlet state $\frac{1}{\sqrt{2}}(|\Phi_+\rangle_1\otimes |\Phi_-\rangle_2 - |\Phi_-\rangle_1\otimes |\Phi_+\rangle_2)$ for $\theta=\pi^-$. Here, $\pi^-$ refers to $\pi-\epsilon$ with $\epsilon$ corresponding to any small deviation from $\theta=\pi$.

The formation of this EPR pair is located around $\theta=\pi$ for the model (\ref{model}) such that we can acquire a good (better) understanding from the poles, generalizing Eqs. (\ref{marker}).

If we measure $A_{j\phi}$ for $d-M<r<d+M$, locally on a sphere at $\theta=\pi$ such that $A_{1\phi}=-i\langle \psi(\pi)| \partial_{1\phi}\otimes\mathbb{I}|\psi(\pi)\rangle$ and $A_{2\phi}=-i\langle \psi(\pi)| \mathbb{I}\otimes \partial_{2\phi}\ |\psi(\pi)\rangle$, then we obtain 
\begin{equation}
\label{identity}
A_{j\phi}(\pi) = \frac{1}{2}A_{j\phi}(0) + \frac{1}{2}A_{j\phi,r=0}(\pi),
\end{equation}
the subscript $r=0$ in the last term referring to the state at south pole $|\psi(\pi)\rangle_{r=0}=|\psi(\pi)\rangle_{r<d-M}$ such that $A_{j\phi,r=0}(\pi)=A_{j\phi,r<d-M}(\pi)$. To write down Eq. (\ref{identity}), we implicitly assume that the choice of the Hilbert space $|\Phi_{\pm}(\phi)\rangle_j$ remains identical
when tuning smoothly the parameters in the Hamiltonian and in particular the interaction $r$. Now, taking into account that for $r=0$ a charge resides inside each sphere we also have 
\begin{equation}
\label{charge}
A_{j\phi,r=0}(\pi) - A_{j\phi}(0) = q.
\end{equation}
Inserting the identity $A_{j\phi,r=0}(\pi) = A_{j\phi}(0)+q$ into Eq. (\ref{identity}) then leads to 
\begin{eqnarray}
C_j &=& A_{j\phi}(\pi) - A_{j\phi}(0) \\ \nonumber
&=& \frac{1}{2\pi}\int_0^{\pi}\int_0^{2\pi} F^j_{\theta\phi}d \theta d\phi =q\frac{1}{2}.
\end{eqnarray} 
Since for the specific class of models in Eq. (\ref{model}), the function $A_{j\phi}$ depends on $\theta$ only such that $F^j_{\theta\phi} = \partial_{\theta} A_{j\phi}(\theta)$, this justifies the writing of $A_{j\phi}(\pi) - A_{j\phi}(0)$ as a topological number i.e. the flux produced by the Berry curvature on the surface of the sphere. The function $A_{j\phi}$ is smoothly defined on the whole surface and we can verify numerically when activating $\theta=vt$ in time that the information at the poles is meaningful and stable for all parameters range $d-M<r<d+M$, even when $M\rightarrow 0^+$ implying $r\rightarrow d$. The fractional number $C_j=\frac{q}{2}$ reveals the formation of an entangled wavefunction at one pole through the factor 
$\frac{1}{2}$ in Eq. (\ref{identity}). This leads to a `true' $\frac{1}{2}$ response on the surface similar to the halved flux quantum in a superconductor $\Phi_s=\frac{\Phi_0}{2}$ with $\Phi_0=\frac{h}{e}$ \cite{Imry,Parks} the flux quantum value in the normal phase. The halved value of $C_j$ is similar to have a half-Skyrmion (i.e. a meron) on a sphere. The sub-system $j$ presents a superposition of two coherent geometries, one encircling the monopole and one participating in the entanglement structure. It is as if we have two tori one on top of each other and where half of the surface encircles the hole, the other surface becoming hidden in the entanglement structure \cite{HH}. The Dirac string information, which can be equivalently formulated as two thin handles on each cylinder,  is also similar to a pair of $\pi$ winding numbers \cite{KarynReview}. 

\section{One-Half Topological Number and EPR pair through Light}

The topological number is equivalently written as $C_j = \frac{q}{2}(\langle \sigma_{jz}(0)\rangle-\langle \sigma_{jz}(\pi)\rangle)=\frac{q}{2}(1-0)$, which is equivalent to say that $-\frac{1}{2}\langle \sigma_{jz}(\theta)\rangle = A_{j\phi}(\theta)$ in the symmetric gauge corresponding to $A_{j\phi}(0)=-\frac{q}{2}$ and $A_{j\phi}(\pi)=0$. Here, we introduce the definition $\langle \psi(\theta)| \sigma_{jz}|\psi(\theta)\rangle=\langle \sigma_z(\theta)\rangle$. 
It is then useful to relate correlation functions at $\theta=\pi$ with the fractional topological number. At north pole, 
\begin{equation}
\label{one}
\langle \psi(0)| \sigma_{1z}\sigma_{2z}|\psi(0)\rangle = |o_{++}(0)|^2 = \frac{2C_j}{q} = 1.
\end{equation}
We have identified $\langle \psi(0)| \sigma_{1z}\sigma_{2z}|\psi(0)\rangle = \langle \sigma_{jz}(0)\rangle^2=\langle \sigma_{jz}(0)\rangle$. Below, we take into account
the conservation of the norm of the two-particles wavefunction from north to south leading to $|o_{++}(0)|^2 = |o_{+-}(\pi)|^2 + |o_{-+}(\pi)|^2$, with here $o_{+-}(\theta) = o_{-+}(\theta)$. At south pole,
\begin{eqnarray}
\label{two}
\hskip -0.45cm \langle \psi(\pi)| \sigma_{1z}\sigma_{2z} |\psi(\pi)\rangle = - |o_{++}(0)|^2 = - \frac{2C_j}{q} = -1.
\end{eqnarray}
We also verify $\langle \psi(\pi)| \sigma_{1x}\sigma_{2x}|\psi(\pi)\rangle = |o_{++}(0)|^2 = \frac{2C_j}{q} = 1$ and $\langle \psi(\pi)| \sigma_{1x}\sigma_{2z}|\psi(\pi)\rangle =0$. For the Bell pair, the von Neumann entanglement entropy $S=-\hbox{Tr}\ln \rho_1\ln \rho_1=-\hbox{Tr} \rho_2\ln \rho_2$, with $\rho_{j}$ corresponding to the reduced density matrix of one spin $j$, can be defined as a measure of entanglement revealing the probability for each spin to be in $|+\rangle_z$ or $|-\rangle_z$ state. At $\theta=0$ and $\theta=\pi$, it takes the simple form $S=-p_+\ln p_+ - p_-\ln p_-$ with $p_{\pm} = \frac{1}{2}(1\pm \langle \sigma_{jz}\rangle)$. For the EPR pair, $p_+=p_-=\frac{1}{2}$ such that we reach a maximum $S(\theta=\pi)=\ln 2$. At north pole, $p_+=1$ and $p_-=0$ leading to $S(\theta=0)=0$. The entanglement entropy is accessible in quantum circuits \cite{entropycircuits}. We have introduced bipartite fluctuations $F$ \cite{bipartitefluctuations} as a (possible) measure of entanglement in many-body quantum systems motivated by an article from J. Bell in 1963 speaking about fluctuations and compressibility theorems in superconductors \cite{Bell1963}. For the present situation, we can build a simple relation between entanglement properties, correlation functions and geometrical responses revealing the same information as the probabilities $p_{\pm}$ in the entropy. In the present situation, the bipartite fluctuations $F=F_1=F_2$ simply correspond to the variance on the measure of a spin magnetization 
\begin{equation}
F(\pi) = \langle \psi(\pi)| \sigma_{1z}^2 \otimes \mathbb{I} |\psi(\pi)\rangle - \langle \psi(\pi)| \sigma_{1z} \otimes \mathbb{I} |\psi(\pi)\rangle^2.
\end{equation}
We verify that
\begin{eqnarray}
\label{three}
\hskip -0.5cm F(\pi) = 4|o_{+-}(\pi)|^2 |o_{-+}(\pi)|^2 = \frac{2C_j}{q} =+1,
\end{eqnarray}
implying $C_j=\frac{1}{2}$ if $q=1$. The maximum values of $F$ and $S$ at the south pole reveal the formation of the EPR pair and also give a clear interpretation to the fractional topological number. At $\theta=0$, $F=S=0$.
Eqs. (\ref{one}), (\ref{two}) and (\ref{three}) are in agreement with the same value $C_j=q\frac{1}{2}$. When $r<d-M$ within the topological phase (such that $M<d$) then we have $\langle \psi(\pi)| \sigma_{1z}\sigma_{2z} |\psi(\pi)\rangle = |o_{++}(0)|^2=\frac{2C_j}{q}$ with $C_j=q$ and $F(\pi)=1-|o_{++}(0)|^2=0$.

We can now propose a measure of the topological number(s) through circularly polarized light.
We can implement a boost of the azimuthal angle $\phi=\mp \omega t$ on one sphere only e.g. through the perturbation $\delta H_1=(-\omega_0 e^{\pm i\omega t}\sigma_1^+ +h.c.)$. At $\theta=\pi$ this is equivalent to introduce a ${\bf d}_1$-vector on sphere $1$ such that $(\tilde{d}\sin{\theta}'\cos\phi, \tilde{d}\sin{\theta}'\sin\phi, \tilde{d}\cos{\theta}'+M)=(\omega_0\cos\omega t,\omega_0 \sin(\mp \omega t), -d+M)$. This leads to the identification $\phi=\omega t$, $\omega_0 = \tilde{d}\sin{\theta}'$ and $\tilde{d}\cos{\theta}'=-d$ for sphere $1$. If $\omega_0\ll d\sim \tilde{d}$ then the angle ${\theta}'\sim\theta$ remains very close to $\pi$ such that $C_j=q\frac{1}{2}$ remain identical for both spins. Activating the azimuthal angle of sphere $1$ through $\phi=\mp\omega t$, for $\phi-\phi'=\mp \frac{\pi}{2}$, from Eq. (\ref{metric}) we also obtain the relation 
\begin{equation}
|_{\phi\phi}\langle \psi(\pi)|\psi(\pi)\rangle_{\phi'\phi}|^2 =\frac{C_j}{q} = \frac{1}{2},
\end{equation}
within the fractional phase which provides another measure of the $\frac{1}{2}$-number. The lowerscript symbol $\phi,\phi$ or $\phi',\phi$ refers to the azimuthal angle of sphere $1$ and $2$, respectively. We obtain a similar result flipping the role of spheres $1$ and $2$. This result is also equivalent as if we modify $\theta=\pi\rightarrow \tilde{\theta}=\frac{\pi}{2}$ in Eq. (\ref{metric}). For one sphere with $r=0$, if we set $\theta=\pi$ and $\phi'-\phi=\pm \frac{\pi}{2}$ in Eq. (\ref{metric}), the same protocol would measure $\alpha_j(\pi)$, i.e. the square of the integer topological invariant \cite{KLHlight}. The measure at north pole reveals an identical response for $r=0$ and for the fractional topological phase, $\alpha_j(0)=\alpha_j(\pi)=1$. 

\section{Cylinder Geometry and Protected Transport}

We can also formulate an analogy with a cylinder geometry, similar to the Laughlin analysis \cite{Laughlin}, where we fix $A_{j\phi}$ at the top and bottom disks to $A_{j\phi}(0)$ and $A_{j\phi}(\pi)$, respectively. Since the function $A_{j\phi}$ and $F^j(\phi,z)=-\partial_z A_{j\phi}$ can be defined smoothly on the vertical surface of the cylinder $\Sigma$ for the model (\ref{model}) then from Stokes' theorem 
\begin{eqnarray}
\label{transportcylinder}
\hskip -0.7cm \frac{1}{2\pi}\int_0^{2\pi}\int_0^{{\cal H}} F^j(\phi,z) d\phi dz &=& (A_{j\phi}(\pi) - A_{j\phi}(0)) = C_j.
\end{eqnarray}
Here, ${\bf F}^j = F^j(\phi,z) {\bf u}$ with ${\bf u}$ a normal unit vector perpendicular to the vertical surface $\Sigma$. This relation is also applicable for the fractional phase. We have the identification $z=\cos\theta$ from the unit sphere. For one sphere when $r=0$ (and $M=0$), $A_{j\phi}(\theta)=-\frac{q}{2}\cos\theta = - \frac{q}{2}z$ such that $F^j(\phi,z)=\frac{1}{2}$. To reproduce the topological number $C_j=q$ then this requires to fix the height associated to the $z$ axis to ${\cal H}=2$. We can also apply a voltage drop from north (top, t) to south (bottom, b) such that $(V_t-V_b)={\cal H}E$ with ${\cal H}=2$ and $E$ is the corresponding electric field. This way, we verify the formation of two edge modes on a cylinder with a conductance $G_j=\frac{e^2}{h}C_j$ such that the transverse pumped current measured along the direction of the azimuthal angle satisfies $J^j_{\perp}=\frac{e^2 C_j}{h} {\cal H} E=\frac{e^2}{h} C_j(V_t-V_b)=(I_t-I_b)$ \cite{KLHlight}. Within this protocol, $J_{\perp}^j$ measures the transport of an electron and of a hole moving in opposite directions along $z$ axis producing effectively a quantum Hall response $2\frac{e^2 C_j}{h}$. These formulae are also applicable for the situation $C_j=\frac{1}{2}$. In this case, to measure a particle this requires a projection onto half of the entangled wavefunction (with $\frac{1}{2}$ probability) justifying the occurrence of $C_j=\frac{1}{2}$ in transport properties. 

The protection of the transverse current within the topological phase can be understood from the fact that the quantization of $J^j_{\perp}$ only depends on the value of $A_{j\phi}$ or $\langle \sigma_{jz}\rangle$ at the poles which remain invariant within the same topological phase as long as we respect the inversion symmetry between the two spins-1/2 \cite{KarynReview}. Interestingly, even in the presence of a disorder in the masses $M_i$ the situation remains positive in the sense that disorder can even produce another elongated stable region with a one-half topological number per sphere \cite{Majoranapole}.

Below, we provide an alternative interpretation through Majorana fermions.

\section{Majorana Fermions}

At north pole, each spin is polarized along $z$ direction such that it is then natural to write $\sigma_{1z}=2c^{\dagger}_1 c_1 -1$ with $c^{\dagger}_1 c_1 |\psi(0)\rangle = +|\psi(0)\rangle$. Equivalently, $|\psi(0)\rangle=c^{\dagger}_1|0\rangle$ corresponds to the creation of a particle (fermion), with  $c^{\dagger}_1 c_1=n_1=0$ or $1$ such that $e^{i\pi c^{\dagger}_1 c_1}=1-2n_1$ and $\{c_1,c_1^{\dagger}\}=1$. We have similar definitions for sphere $2$ in terms of the fermion $c^{\dagger}_2|0\rangle$. The model close to $\theta=\pi$ can then be mapped onto four Majorana fermions through the Jordan-Wigner transformation \cite{JordanWigner} such that $\sigma_{1z}=\frac{1}{i}(c^{\dagger}_1-c_1)$, $\sigma_{1x}=(c^{\dagger}_1+c_1)$, $\sigma_{2z}=\frac{1}{i}(c^{\dagger}_2-c_2)e^{i\pi c^{\dagger}_1 c_1}$ and $\sigma_{2x}=(c^{\dagger}_2+c_2)e^{i\pi c^{\dagger}_1 c_1}$ \cite{ArticleThese}. We introduce Majorana fermions which are their own anti-particles  $\eta_j = \frac{1}{\sqrt{2}}(c_j+c_j^{\dagger})=\eta_j^{\dagger}$ and $\alpha_j=\frac{1}{\sqrt{2}i}(c^{\dagger}_j-c_j)=\alpha_j^{\dagger}$, such that $\{\eta_j,\eta_j\}=1=\{\alpha_j,\alpha_j\}=1$ with $2i\eta_j\alpha_j=1-2c^{\dagger}_j c_j$. The choice of `spin-Majorana fermions' representation here takes into account the formation of the EPR pair leading to $\langle \psi(\pi^-)|\sigma_{jz}| \psi(\pi^-)\rangle=0$ within the ground state. The effective Hamiltonian is written as
\begin{eqnarray}
H_{eff} &=& r\sigma_{1z}\sigma_{2z} - \frac{d^2\sin^2\theta}{r}\sigma_{1x}\sigma_{2x} \\ \nonumber
&=& -2r i \eta_1\alpha_2 - \frac{2i d^2}{r}\sin^2\theta\alpha_1\eta_2.
\end{eqnarray}
The ground state satisfies $\langle \psi(\pi)| \sigma_{1z}\sigma_{2z} |\psi(\pi)\rangle = \langle \psi(\pi)| 2i \alpha_2 \eta_1 |\psi(\pi)\rangle = -1$. At $\theta=\pi$, the system shows two free Majorana fermions 
$\alpha_1$ and $\eta_2$ encoding the degeneracy of $|\Phi_+\rangle_1|\Phi_-\rangle_2$ and $|\Phi_-\rangle_1|\Phi_+\rangle_2$, which become bounded at $\theta=\pi^-$ due to the term $-\frac{d^2\sin^2\theta}{r}$ such that $\langle \psi(\pi^-)| 2i \alpha_1 \eta_2 |\psi(\pi^-)\rangle = 1$. This implies $\langle \psi(\pi^-)| c^{\dagger}_j c_j |\psi(\pi^-)\rangle =\frac{1}{2}$, such that one fermion at $\theta=0$ has a probability $\frac{1}{2}=C_j$ to reach $\theta=\pi$. We also have $\langle \psi(\pi)|c^{\dagger}_j c_j |\psi(\pi)\rangle =\frac{1}{2}$ revealing the zero-energy (free) Majorana fermions through $\langle \psi(\pi)| 2 i \eta_j \alpha_j |\psi(\pi)\rangle=0$ \cite{comment}. A pair of $1/2$-topological numbers can then refer to a pair of free Majorana fermions at one pole. This identification is then generalizable to other systems such as interacting superconducting Kitaev wires \cite{pwavewire}.

Here, we show that the local characterization of the topological properties on the sphere and the occurrence of an EPR pair at one pole can also be applied as a marker in topological energy band structures.

\section{Application into Band Theory}

The two spheres' model can be realized with two planes $1$ and $2$. The radial magnetic field on each sphere corresponds to a Haldane model \cite{Haldanemodel}, where the two spin polarizations' eigenstates $|\Phi_{\pm}\rangle_j$ refer to the occupancy on sub-lattice $A$ or $B$ of the honeycomb lattice for a plane $j$ \cite{HH}. The parameter $r$ corresponds to a hopping term between planes (in a AA-BB stacking, which can be e.g. realized in optical lattices \cite{Bilayer2planes}) and $M$ describes a staggered potential on the lattice. For $r<d-M$, each plane forms a quantum anomalous Hall phase with a topological number $|C_j|=1$. For $r>d+M$, each plane is an insulator characterized by a topological number $C_j=0$. The two lowest bands are characterized through opposite topological numbers \cite{Bilayer2planes}. Tuning $r$ such that $d-M<r<d+M$, there is the formation of a nodal ring semimetal at the Fermi energy \cite{HH}. The Dirac points $K$ and $K'$ within the Brillouin zone correspond to the north and south poles respectively. 

The two-particles wavefunction at the $K$ point becomes identical to the one on the sphere at $\theta=0$, flipping the role of the two-spin polarizations as a result of the inversion ${\bf d}_i\rightarrow -{\bf d}_i$ in the definition of the Hamiltonian: $|\psi(K)\rangle = |\psi_1(K)\rangle\otimes|\psi_2(K)\rangle=e^{i\pi} c^{\dagger}_{B1}c^{\dagger}_{B2}|0\rangle_K=e^{i\pi}|\Phi_-\rangle_1^1\otimes|\Phi_-\rangle_2^2$. The state $|\psi_1(K)\rangle\otimes |\psi_2(K)\rangle$ associated to two particles occupying bands $1$ and $2$, at wave-vector $K$, corresponds to two particles polarized on sub-lattice $B$ occupying a different plane (or sphere). In the definition of $|\Phi_{\pm}\rangle_j^k$, the lowerscript $j=1$ or $2$ refers to sphere or plane index and the superscript to particle $k=1$ or $2$. We can measure geometrical properties
in each plane locally resolved at the $K$ point such that 
\begin{equation}
A_{j\phi}(K)=-i\langle \psi(K)| \partial_{j\phi} |\psi(K)\rangle = A_{j\phi}(0)
\end{equation}
where $A_{j\phi}(0)$ on the right-hand side refers to the geometrical response at north pole on a sphere. The index $j$ on the left-hand side refers to a measure  
on a plane. The two-particles' wavefunction is uniquely defined around $K$ justifying the smoothness of $A_{j\phi}(K)$ in this area. 

At the $K'$ point, the ground-state is also non-degenerate $|\psi(K')\rangle=|\psi_1(K')\rangle\otimes |\psi_2(K')\rangle$ (with $1$ and $2$ corresponding to the two occupied energy bands). This can also be written as $|\psi(K')\rangle = \frac{1}{2}(-c^{\dagger}_{A1} +c^{\dagger}_{A2})(-c^{\dagger}_{B1}+c^{\dagger}_{B2})|0\rangle_{K'}$ in the formulation of the two planes. This represents an entangled wavefunction where each particle is delocalized between the two planes and has an equal probability $\frac{1}{2}$ to be in each plane. Below, we discuss several ways to interpret this probability $\frac{1}{2}$. 

In the spheres (planes) and particles representation, 
\begin{eqnarray}
\label{wavefunctionK'}
|\psi(K')\rangle &=& \frac{1}{2}e^{i\pi}(|\Phi_+\rangle_1^1\otimes|\Phi_-\rangle_2^2 + |\Phi_-\rangle_1^1\otimes|\Phi_+\rangle_2^2)\\ \nonumber
&+& \frac{1}{2}(|\Phi_+\rangle_1^1\otimes|\Phi_-\rangle_1^2 + |\Phi_-\rangle_2^1\otimes|\Phi_+\rangle_2^2).
\end{eqnarray}
We have modified $|\Phi_+\rangle_2^2\rightarrow -|\Phi_+\rangle_2^2$ such that the first line corresponds to $|\psi(\pi^-)\rangle$, from the triplet sector, on the sphere. There is one particle on one sphere (plane) for this state. The second line reveals two particles on one sphere (plane). 
Evaluating $-i\langle \psi(K')| \partial_{j\phi} | \psi(K')\rangle$ where $j$ corresponds to a measure on a plane $j$ then we have
\begin{equation}
\label{equationK'}
A_{j\phi}(K') = 2\frac{1}{4}(A_{j\phi}(0) + A_{j\phi,r=0}(\pi))= A_{j\phi}(\pi).
\end{equation}
The function $A_{j\phi}(K')$ is smooth around $K'$ with the information resolved in each plane. To show this equality, we have introduced $A_{j\phi}(0)$ and $A_{j\phi,r=0}(\pi)$ referring to the two-spheres model in Eq. (\ref{model}) leading to Eq. (\ref{identity}). 
Due to the symmetric form of Eq. (\ref{wavefunctionK'}), we obtain the same result if we invert the role of planes and particles.

Measuring $A_{j\phi}(K)$ and $A_{j\phi}(K')$ around each Dirac point, $A_{j\phi}(K')-A_{j\phi}(K)$ can be defined as a local topological marker related to the fractional phase of the two-spheres' model. We can then write down $A_{j\phi}(K)$ in terms of $-\frac{1}{2}\langle\sigma_{jz}(0)\rangle$ and $A_{j\phi}(K')=-\frac{1}{2}\langle\sigma_{jz}(\pi)\rangle$, such that $A_{j\phi}(K')-A_{j\phi}(K) = \frac{1}{2}(\langle \sigma_{jz}(0)\rangle -\langle \sigma_{jz}(\pi)\rangle)=q\frac{1}{2}$. Here, $\langle\sigma_{jz}\rangle$ measures the relative occupancies on sublattices $A$ and $B$ resolved at each Dirac point in a plane $j$. It is also in agreement with the structure of the edge mode which shows a 50\%-50\% probability to occupy a plane \cite{HH}. In this sense $q\frac{1}{2}$ can be introduced as a topological bulk quantity defined on each side of the band-crossing region at the Fermi energy, from the two-spheres' model within the fractional phase. This results in a total Berry phase $2\pi(A_{j\phi}(K')-A_{j\phi}(K))=q\pi$, with $q=-1$, resolved at the two Dirac points \cite{comment2}.

Now, we elaborate on the relation between this local topological marker and the quantum Hall responses. We introduce the angle $\theta_c$ on the sphere that will precisely refer to the location of the band-crossing for the topological semimetal along the path in the Brillouin zone. For the model in Eq. (\ref{model}), from Stokes' theorem, $C_j=A_{j\phi}(\pi)-A_{j\phi}(0)=A'_{j\phi}(\theta<\theta_c)-A'_{j\phi}(\theta>\theta_c)=q\frac{1}{2}$ with $A_{j\phi}(\theta)$ being smoothly defined for $\theta\in[0;\pi]$ \cite{HH}. We introduce $A'_{j\phi}(\theta<\theta_c)=A_{j\phi}(\theta_c)-A_{j\phi}(0)=\int_0^{\theta_c}F_{\theta \phi}^j d\theta$ and $A'_{j\phi}(\theta>\theta_c)=A_{j\phi}(\theta_c)-A_{j\phi}(\pi)=-\int_{\theta_c}^{\pi} F^j_{\theta\phi} d\theta$. The functions $A'_{j\phi}(\theta<\theta_c)$ and $A'_{j\phi}(\theta>\theta_c)$ characterize the transverse currents $J_{\perp}^e(\theta<\theta_c)$, for an electron going from $\theta=0$ to $\theta=\theta_c$, and $J_{\perp}^{h}(\theta>\theta_c)$ for a hole going from $\theta=\theta_c$ to $\theta=\pi$ within the same sphere. Therefore, the local marker $A_{j\phi}(K')-A_{j\phi}(K)$ measures the quantum Hall conductivity $|q|\frac{1}{2}\frac{e^2}{h}$  related to the model in Eq. (\ref{model}). We can also elucidate the information within the two occupied bands of the topological semimetal from the momentum space. If we integrate the Berry curvature on the same two domains $\theta\in [0;\theta_c^-[$ and $\theta\in ]\theta_c^+;\pi]$, and sum the contributions of the two occupied bands on each domain, this gives rise to an edge contribution on a cylinder associated to a plane of the form $A_{j\phi}(K') - A_{j\phi}(K)=q\frac{1}{2}$, related to the local topological marker. This can also be interpreted as a half quantized `conductivity' response in units of $e^2/h$. Now, one may question the role of the crossing region. This gives rise to an additional `bulk' contribution located symmetrically around the crossing point (on the cylinder in momentum space) of the form $A_{j\phi,r=0}(\tilde{\theta}_c)$ per plane if we introduce the dressed angle such that $\tan\tilde{\theta}_c=\sin \theta_c/(\cos \theta_c +\frac{M}{d})$. This bulk contribution is zero (per plane) if we define a half-Skyrmion on the sphere $\tilde{\theta}_c=\frac{\pi}{2}$, which requires to satisfy $\cos\theta_c+\frac{M}{d}=0$ with $d-M<r<d+M$. In this case, the half-quantized Hall response in each plane comes from the region 
$\tilde{\theta}\in [0;\tilde{\theta}_c=\frac{\pi}{2}]$ on a sphere, revealing the topological formation of a half-Skyrmion. 

The topological nodal ring semimetal can also be realized in one graphene plane \cite{oneplanesemimetal}. For this situation, for band $2$, particles at $\theta_c^{\pm}$ acquire orthogonal spin polarizations $|+\rangle_x$ and $|-\rangle_x$ \cite{Sariahtransport}, and the energy band $2$ can develop a quantum Hall \cite{TKNN} and/or a quantum spin Hall response \cite{KaneMele}. Interestingly, the responses to circularly polarized light locally at the Dirac points can measure signatures of the one-half topological number related to the quantized quantum Hall response of the lowest energy band \cite{KarynReview,Sariahtransport}.

\section{Summary}

To summarize, we have shown how the formation of an EPR pair or Bell state locally on the Bloch sphere can be characterized through half-topological numbers and zero-energy Majorana fermions. The half-quantized response can be revealed through a circularly polarized field and also through transport. We have shown the relevance of the two-spheres' model to characterize the (topological) edge response in bilayer geometries. These spheres' models can be realized \cite{Roushan} and may find applications in energy through a dynamo effect \cite{dynamo1,dynamo2}. The presence of zero-energy Majorana fermions can find relevance for quantum information \cite{Majoranapole}. The physics of two entangled spins-1/2 may also have relations towards black hole physics, e.g. through similarities 
 between EPR pairs and the formation of ER bridges \cite{Maldacena}. It is also interesting to mention recent applications of one-half magnetic monopoles through Berry phases \cite{DeguchiFujikawa}.

This work has benefitted from discussions with J. Hutchinson, S. Al Saati, E. Bernhardt, F. del Pozo and J. Legendre. We acknowledge discussions at Aspen Center for Physics, January 2022, conference on New Directions in Strong Correlation Physics, at UCLondon July 2023, and when giving 4 lectures on geometry and topology in the quantum at Paris-Saclay lectures June 2023. This work was supported by the french ANR BOCA and the Deutsche Forschungsgemeinschaft (DFG), German Research Foundation under Project No. 277974659.


\begin{thebibliography}{0}%
\makeatletter
\providecommand \@ifxundefined [1]{%
 \@ifx{#1\undefined}
}%
\providecommand \@ifnum [1]{%
 \ifnum #1\expandafter \@firstoftwo
 \else \expandafter \@secondoftwo
 \fi
}%
\providecommand \@ifx [1]{%
 \ifx #1\expandafter \@firstoftwo
 \else \expandafter \@secondoftwo
 \fi
}%
\providecommand \natexlab [1]{#1}%
\providecommand \enquote  [1]{``#1''}%
\providecommand \bibnamefont  [1]{#1}%
\providecommand \bibfnamefont [1]{#1}%
\providecommand \citenamefont [1]{#1}%
\providecommand \href@noop [0]{\@secondoftwo}%
\providecommand \href [0]{\begingroup \@sanitize@url \@href}%
\providecommand \@href[1]{\@@startlink{#1}\@@href}%
\providecommand \@@href[1]{\endgroup#1\@@endlink}%
\providecommand \@sanitize@url [0]{\catcode `\\12\catcode `\$12\catcode
  `\&12\catcode `\#12\catcode `\^12\catcode `\_12\catcode `\%12\relax}%
\providecommand \@@startlink[1]{}%
\providecommand \@@endlink[0]{}%
\providecommand \url  [0]{\begingroup\@sanitize@url \@url }%
\providecommand \@url [1]{\endgroup\@href {#1}{\urlprefix }}%
\providecommand \urlprefix  [0]{URL }%
\providecommand \Eprint [0]{\href }%
\providecommand \doibase [0]{https://doi.org/}%
\providecommand \selectlanguage [0]{\@gobble}%
\providecommand \bibinfo  [0]{\@secondoftwo}%
\providecommand \bibfield  [0]{\@secondoftwo}%
\providecommand \translation [1]{[#1]}%
\providecommand \BibitemOpen [0]{}%
\providecommand \bibitemStop [0]{}%
\providecommand \bibitemNoStop [0]{.\EOS\space}%
\providecommand \EOS [0]{\spacefactor3000\relax}%
\providecommand \BibitemShut  [1]{\csname bibitem#1\endcsname}%
\let\auto@bib@innerbib\@empty
\end{thebibliography}%


\begin{thebibliography}{9}

\bibitem{QiZhang}
X. Liang Qi and S.-C. Zhang, Topological insulators and superconductors, Rev. Mod. Phys. {\bf 83}, 1057 (2011). J. Maciejko, T. L. Hughes and S.-C. Zhang, Annual Review of Condensed Matter Physics. {\bf 2}: 31–53 (2011).
B. A. Bernevig with T. Hughes,  Topological insulators and Topological superconductors, Princeton University Press (2013).

\bibitem{TIreview}
Z. Hasan and C. L. Kane, Colloquium: Topological insulators, Rev. Mod. Phys. {\bf 82}, 3045 (2010).

\bibitem{QH}
K. Von Klitzing, G. Dorda and M. Pepper, New method for high accuracy determination of the fine structure constant based on quantized Hall resistance, Phys. Rev. Lett. {\bf 45}, 494 (1980).

\bibitem{HH}
J. Hutchinson and K. Le Hur, Quantum entangled fractional topology and curvatures, arXiv:2002.11823, Communications Physics {\bf 4}, 144 (2021).

\bibitem{Roushan}
P. Roushan, C. Neill, Yu Chen, M. Kolodrubetz, C. Quintana, N. Leung, M. Fang, R. Barends, B. Campbell, Z. Chen, B. Chiaro, A. Dunsworth, E. Jeffrey, J. Kelly, A. Megrant, J. Mutus, P. J. J . O'Malley, D. Sank, A. Vainsencher, J. Wenner, T. White, A.  Polkovnikov, 
A. N. Cleland and J. M. Martinis, Observation of topological transitions in interacting quantum circuits, Nature {\bf 515}, 241-244 (2014).

\bibitem{Boulder}
M. D. Schroer, M. H. Kolodrubetz, W. F. Kindel, M. Sandberg, J. Gao, M. R. Vissers, D. P. Pappas, Anatoli Polkovnikov and K. W. Lehnert, Measuring a Topological Transition in an Artificial Spin-System, Phys. Rev. Lett. {\bf 113}, 050402 (2014).

\bibitem{Berry}
M. V. Berry, Quantal Phase Factors Accompanying Adiabatic Changes, Proceedings of the Royal Society of London. Series A, Mathematical and Physical Sciences {\bf 392}, Issue 1802, Page 45 (1984).

\bibitem{Skyrmion1}
S. L. Zhang, G. van der Laan and T. Hesjedal, Direct experimental determination of the topological winding number of skyrmions in Cu$_2$OSeO$_3$, Nature Communications {\bf 8}, 14619 (2017).

\bibitem{Skyrmion2}
T. H. R. Skyrme, A Unified Field Theory of Mesons and Baryons, Nuclear Physics {\bf 31}, 556-569 (1962). 

\bibitem{KLHlight}
K. Le Hur, Global and local topological quantized responses from geometry, light, and time, Phys. Rev. B {\bf 105}, 125106  (2022).

\bibitem{dynamo1}
L. Henriet, A. Slocchi, P. P. Orth and K. Le Hur, Topology of a dissipative spin: dynamical Chern number, bath induced non-adiabaticity and a quantum dynamo effect, Phys. Rev. B {\bf 95}, 054307 (2017).

\bibitem{dynamo2}
E. Bernhardt, C. Elouard and K. Le Hur, A topologically protected quantum dynamo effect in a driven spin-boson model, Phys. Rev. A {\bf 107}, 022219 (2023).

\bibitem{AlainAspect}
A. Aspect, Bell's Theorem : The Naive View of an Experimentalist, text prepared for a talk at a conference in memory of John Bell, held in Vienna in December 2000. Published in ``Quantum [Un]speakables - From Bell to Quantum information'', edited by R. A. Bertlmann and A. Zeilinger, Springer (2002); arXiv:quant-ph/0402001.

\bibitem{bipartitefluctuations}
H. F. Song, S. Rachel, C. Flindt, I. Klich, N. Laflorencie and K. Le Hur, Bipartite fluctuations as a probe of many-body entanglement, Phys. Rev. B {\bf 85}, 035409 (2012).

\bibitem{Majorana}
E. Majorana, Teoria simmetrica dell’elettrone e del positrone, Il Nuovo Cimento, {\bf 14}, 171-184 (1937).

\bibitem{Majoranareturns}
F. Wilczek, Majorana Returns, Nature Physics {\bf 5}, 614-618 (2009).

\bibitem{Actor}
A. Actor, Classical solutions of SU(2) Yang-Mills theories, Rev. Mod. Phys. {\bf 51}, 461 (1979).

\bibitem{Sondhi}
S. L. Sondhi, A. Karlhede, S. A. Kivelson, and E. H. Rezayi, Skyrmions and the crossover from the integer to fractional quantum Hall effect at small Zeeman energies, Phys. Rev. B {\bf 47}, 16419 (1993).

\bibitem{Shayegan}
Y. P. Shkolnikov, S. Misra, N. C. Bishop, E. P. De Poortere and M. Shayegan, Observation of Quantum Hall Valley Skyrmions, Phys. Rev. Lett. {\bf 95}, 066809 (2005).

\bibitem{Potemski}
L. Bryja, K. Ryczko, A. W\' ojs, J. Misiewicz and M. Potemski, Skyrmions in a Hole Gas with Large Spin Gap and Strong Disorder, Acta Physica Polonica A, {\bf 110}, 2 (2006).

\bibitem{Xu}
Y. Xu, I. Miotkowski, C. Liu,  J. Tian, H. Nam, N. Alidoust, J. Hu, C.-K. Shih, M. Z. Hasan and Y. P. Chen,  Observation of topological surface state quantum Hall effect in an intrinsic three-dimensional topological insulator, Nature Phys {\bf 10}, 956-963 (2014).

\bibitem{Jena}
J. Jena, B. G\" obel, T. Hirosawa, S. A. Díaz, D. Wolf, T. Hinokihara, V.  Kumar, I. Mertig, C. Felser, A. Lubk, D. Loss and S. S. P. Parkin, Observation of fractional spin textures in a Heusler material, Nature Communications {\bf 13}, 2348 (2022).

\bibitem{Fert}
A. Fert, N. Reyren and V. Cros, Advances in the Physics of Magnetic Skyrmions and Perspective for Technology, Nat Rev Mater {\bf 2}, 17031 (2017).

\bibitem{Nagaosa}
N. Nagaosa and Y. Tokura, Topological properties and dynamics of magnetic skyrmions, Nature Nanotech {\bf 8}, 899-911 (2013).

\bibitem{Carollo}
A. Carollo, D. Valenti and B. Spagnolo, Geometry of quantum phase transitions, Physics Reports {\bf 838}, 1-72 (2020).

\bibitem{Tan}
X. Tan, D.-W. Zhang, Z. Yang, Ji Chu, Y.-Q. Zhu, D. Li, X. Yang, S. Song, Z. Han, Z. Li, Y. Dong, H.-F. Yu, H. Yan, S.-L. Zhu, and Y. Yu, Experimental Measurement of the Quantum Metric Tensor and Related Topological Phase Transition with a Superconducting Qubit,
 Phys. Rev. Lett. {\bf 122}, 210401 (2019).

\bibitem{Haldanemodel}
F. D. M. Haldane, Model for a Quantum Hall Effect without Landau Levels: Condensed-Matter Realization of the "Parity Anomaly", Phys. Rev. Lett. {\bf 61}, 2015 (1988).

\bibitem{Klein}
Ph. W. Klein, A. Grushin and K. Le Hur, Interacting stochastic topology and Mott transition from light response, Phys. Rev. B {\bf 103}, 035114 (2021).

\bibitem{pwavewire}
F. del Pozo, L. Herviou, K. Le Hur, Fractional topology in interacting one-dimensional superconductors, Phys. Rev. B {\bf 107}, 155134 (2023).

\bibitem{Imry}
Y. Imry, Introduction to Mesoscopic Physics, Oxford University Press, 1997, New York.

\bibitem{Parks}
R. D. Parks, Quantized Magnetic Flux in Superconductors: Experiments confirm Fritz London's early concept that superconductivity is a macroscopic quantum phenomenon, Science {\bf 146}, Issue 3650, 1429-1435 (1964).

\bibitem{KarynReview}
K. Le Hur, Topological Matter and Fractional Entangled Quantum Geometry through Light, arXiv:2209.15381, review submitted to Physics Reports 109 pages, Elsevier, September 2023.

\bibitem{entropycircuits}
    C. Neill, P. Roushan, M. Fang, Y. Chen, M. Kolodrubetz, Z. Chen, A. Megrant, R. Barends, B. Campbell, B. Chiaro, A. Dunsworth, E. Jeffrey, J. Kelly, J. Mutus, P. J. J. O’Malley, C. Quintana, D. Sank, A. Vainsencher, J. Wenner, T. C. White, A. Polkovnikov and J. M. Martinis, Ergodic dynamics and thermalization in an isolated quantum system, Nature Physics  {\bf 12}, pages 1037-1041 (2016).     
   
\bibitem{Bell1963}
J. S. Bell, Fluctuation Compressibility Theorem and Its Application to the Pairing Model, Phys. Rev. {\bf 129}, 1896 (1963).

\bibitem{Laughlin}
R. B. Laughlin, Quantized Hall conductivity in two dimensions, Phys. Rev. B {\bf 23}, 5632(R) (1981).

\bibitem{Majoranapole}
E. Bernhardt, B. C. H. Cheung and K. Le Hur, Majorana fermions and quantum information with fractional topology and disorder, arXiv:2309.03127.

\bibitem{JordanWigner}
P. Jordan and E. Wigner, \"Uber das Paulische \"Aquivalenzverbot, Z. Physik {\bf 47}, 631-651 (1928).

\bibitem{ArticleThese}
K. Le Hur and B. Coqblin, Underscreened Kondo effect: A two S=1 impurity model, Phys. Rev. B 56, 668 (1997).

\bibitem{comment}
When navigating from north to south, the topological number is precisely measured at $\theta=\pi^-$ \cite{HH,KarynReview}. This also requires the smoothness of the functions $A_{j\phi}$ and $\langle \psi| \sigma_{jz} |\psi\rangle$ which is implicitly
verified through these identities since $\langle \psi(\pi)| \alpha_1 | \psi(\pi) \rangle = \langle \psi(\pi^-)| \alpha_1 | \psi(\pi^-) \rangle=0$.  

\bibitem{Bilayer2planes}
P. Cheng, P. W. Klein, K. Plekhanov, K. Sengstock, M. Aidelsburger, C. Weitenberg and K. Le Hur, Topological proximity effects in a Haldane graphene bilayer system, Phys. Rev. B {\bf 100}, 081107(R) (2019).

\bibitem{comment2}
On the sphere, $A_{j\phi}(K')-A_{j\phi}(K)=q\pi$ corresponds indeed to the addition of Berry phases due to a modification of $\phi\rightarrow - \phi$ at south pole \cite{KarynReview}.

\bibitem{oneplanesemimetal}
K. Le Hur and S. Al Saati,  Topological nodal ring semimetal in graphene, Phys. Rev. B {\bf 107}, 165407 (2023).

\bibitem{Sariahtransport}
K. Le Hur and S. Al Saati,  Quantum Hall and Light Responses in a 2D Topological Semimetal, arXiv:2311.13922.

\bibitem{TKNN}
D. Thouless, M. Kohmoto, M. P. Nightingale and M. den Nijs, Quantized Hall Conductance in a Two-Dimensional Periodic Potential, Phys. Rev. Lett. {\bf 49}, 405 (1982).

\bibitem{KaneMele}
C. L. Kane and E. Mele, Quantum Spin Hall Effect in Graphene, Phys. Rev. Lett. {\bf 95}, 226801 (2005); C. L. Kane and E. Mele, Z$_2$ Topological Order and the Quantum Spin Hall Effect, Phys. Rev. Lett. {\bf 95}, 146802 (2005).

\bibitem{Maldacena}
J. Maldacena and L. Susskind, Cool horizons for entangled black holes, Progress of Physics {\bf 61}, Issue 9, 781-811 (2013).

\bibitem{DeguchiFujikawa}
S. Deguchi and K. Fujikawa, A new magnetic monopole inspired by Berry’s phase, Physics Letters B {\bf 802}, 135210 (2020). 


\end{thebibliography}
\end{document}